\documentstyle{lamuphys}
\makeatletter
\let\chapter\hid@chapter
\makeatother
\begin{document}
\pagenumbering{arabic}
\title{Large-Scale Structure at High Redshift}

\author{Simon\,D.M.\,White}

\institute{Max-Planck-Institut f\"ur Astrophysik,
Karl-Schwarzschild-Stra{\ss}e 1,\\ D-85740 Garching bei M\"unchen, Germany}

\maketitle

\begin{abstract}
I discuss and illustrate the development of large-scale structure in the 
Universe, emphasising in particular the physical processes and cosmological 
parameters that most influence the observationally accessible
aspects of structure at large redshift. Statistical properties of this
structure can be measured from the apparent positions of faint
galaxies and quasars; the structure can be mapped in three 
dimensions by obtaining redshifts for large samples of such objects;
it can be studied using foreground absorption in the spectra of
quasars; finally the 
mass distribution can be constrained by measuring the
gravitationally induced distortion of background galaxy images. The
first and last of these techniques require deep imaging of large areas
of the sky with the best possible image quality. The second and third
will require 8m-class telescopes with efficient multiobject
spectrographs. For QSO absorption line spectroscopy high spectral
resolution is also important.

\end{abstract}
\section{Introduction}
The term ``large-scale structure'' normally refers to the distribution
of matter on scales larger than those of individual galaxies or
galaxy clusters. On these scales objects have not yet had time to
collapse fully and to come to equilibrium, and so their observed
morphology is determined principally by the properties of the initial 
fluctuations from which they grew, and by the physical processes which
amplified those fluctuations. As a result, there is hope that by
studying large-scale structure we may learn directly about the
mechanisms which imprinted irregularities on our otherwise almost
homogeneous Universe.

Recent observational work on large-scale structure has focussed
primarily on galaxy redshift surveys, the most recent
example, and also the largest so far, being the Las Campanas Redshift
Survey of more than 25,000 galaxies (Shectman et al 1996).
The planned 2dF and Sloan surveys will increase this already
impressive number by more than an order of magnitude. Surveys of this
kind can be analysed in many ways but two broad approaches can be
distinguished, quantitative analysis of low-order statistics
and determinations of the morphology and topology of structure through
detailed maps. The first approach typically aims to discriminate
between specific models such as the many variants of the cold dark matter
(CDM) model, while the second is more empirical and might, for
example, provide a test for the broad class of theories which
assume that structure grew through gravitational amplification of an
initially gaussian density fluctuation field. In the nearby
Universe independent distances to galaxies can be measured sufficiently
well for studies of peculiar velocities and large-scale flows to
provide powerful additional constraints on the large-scale mass
distribution.

Theoretical discussions of large-scale structure vary from the
development of purely descriptive statistics for the present
distribution of galaxies (e.g. power spectra, position and velocity
correlations of all orders, counts-in-cells, 
void probabilities and the relations
between these quantities) through dynamical treatments of structure
growth based on perturbation theory, to massive attempts to simulate
the development of structure from the linear into the fully nonlinear
regime. In this contribution I will focus mainly on the latter since
simulations produce results that are easy to appreciate visually and 
can be compared directly with observational data. The descriptive 
statistics do, of course, provide the main quantitative comparison
between simulation and observation, while the perturbation theory
gives a powerful means for checking that the numerical simulations are, in
fact, correct. The main difficulty when comparing
theory and observation is 
that most current simulations predict the distribution of dark matter
whereas almost all the observational data refer to the
distributions of gas or of galaxies. The first attempts
to simulate both gas dynamics and galaxy formation have now been
carried out,  but much of the physics cannot be
treated properly and the results must be regarded as very uncertain.   

In the next section I discuss simulations of structure formation in
somewhat more detail in order to to show how the evolution of
large-scale structure depends on the cosmological model in which it is
occurring. I then summarise our current understanding of the relation
between visible structure and that in the mass
in order to indicate how the predictions of N-body simulations
may be related to observation. Section 4 discusses the techniques
available for measuring large-scale structure at high redshift
and assesses what may be achievable with the
next generation of telescopes if current ideas about structure
formation are correct.

\section{Evolution of Structure in the Dark Matter}

Standard structure formation theories suppose the dark matter
to be pregalactic, collisionless, and gravitationally
dominant. The only significant agent affecting the recent development
of its spatial structure is then gravity, and on large scales the 
distribution of the visible material is also structured
primarily by the gravity of the dark matter. The currently most
popular, and certainly the most thoroughly investigated structure
formation models suppose that the dark matter is non-baryonic
and that the initial deviations from uniformity were produced
by quantum fluctuations during an early inflationary period. The
latter assumption implies that density fluctuations at early times are
a gaussian random field and so are fully specified by their power
spectrum alone. The form of this power spectrum depends on the details
of the inflation model and on the nature of the dark matter. Since
early work showed that hot dark matter, specifically neutrinos
with a mass of a few tens of eV, cannot produce the kind of 
large-scale structure we see (they produce structure too late and on
scales which are too large) inflationary 
models now all assume the dominant matter constituent to be some form
of cold dark matter (CDM). Topological relics of an
early phase transition are another possibility for imposing
structure on an otherwise uniform universe (e.g. Brandenberger 1994); 
I will not discuss them further here.

Within the general family of CDM models, several cures have been
proposed for the inability of the original ``standard'' CDM
model simultaneously to
fit data on large-scale structure and the fluctuation 
amplitude measured by COBE. All involve adding an additional
complexity to the model. Thus ``tilted'' CDM (or TCDM)
supposes that a non-standard inflation model produces 
fluctuations with a slightly different scaling of amplitude with
wavelength; hot plus cold, or mixed dark matter models (H+CDM or MDM)
suppose that that a small fraction of the dark matter is in the form
of stable massive neutrinos; $\tau$CDM supposes that the decay of an
unstable massive neutrino at early times has left a relativistic
neutrino background of higher density than in the standard model;
$\Lambda$CDM supposes that a cosmological constant makes a significant
contribution to the present energy density; open CDM
(OCDM) supposes the curvature radius of the Universe to be comparable
to its observable extent, rather than much larger as predicted by
standard inflation. With a suitable choice of the additional
free parameter each of these models can be made to give a rough fit
both to the COBE amplitude and to observed large-scale structure.

To break the degeneracy between these models, one must
appeal to other data. Large-scale flows may exclude low density
models (Dekel 1995); combining measures of the Hubble constant and 
of globular cluster ages may exclude high density models (Freedman et al 
1994); in a high density universe the observed baryon fraction in galaxy 
clusters may be inconsistent with big bang nucleosynthesis
(White at al. 1993); the Hubble diagram for distant SNIa or the 
frequency of gravitational lensing of quasars may rule out models with a
substantial cosmological constant (Perlmutter et al 1996; Kochanek 1995).
For the purposes of this talk, however, the major difference between
the high and low density models lies in the predicted evolution of
large-scale structure with redshift. 

This difference is
illustrated in figures 1 to 3. These plots show thin slices through
some large N-body simulations of a ``standard'' CDM model (SCDM)  
and of some variants that are generally
thought to be consistent both with COBE and with present-day
large-scale structure. These pictures were made by Joerg Colberg from
simulations carried out on the Garching T3D parallel supercomputer as 
part of the programme of the Virgo Consortium (Jenkins et al, in 
preparation). The
simulations used 17 million particles to follow the evolution of the 
matter distribution within comoving cubic regions of present size 
$240h^{-1}$Mpc; they are able to resolve structures down to a linear
scale of $25h^{-1}$kpc and a mass scale corresponding to the halo of
a Milky Way-like galaxy. The thickness of each slice is about 10\% of its
width. The dark matter distribution is smoothed adaptively to give an
overdensity which is represented on the same logarithmic
colour scale in all plots. Objects containing fewer than 20 particles are not
visible. 

\begin{figure}[t] 
\vspace{14.9cm}
\caption{Slices through simulations of four cosmological
models at $z=0$. 
}
\end{figure}
  
Structure is much more prominent at $z=0$ in the low 
density models (OCDM and $\Lambda$CDM) than in the Einstein-de Sitter 
models ($\tau$CDM and SCDM). This is a reflection of the well-known 
``bias' needed to make high density models consistent with the observed
galaxy distribution. Although all four models have about the 
same abundance of massive quasi-equilibrium objects --
rich galaxy clusters -- models with a high total matter content
achieve this with relatively low fluctuation amplitudes, lower, in 
fact, than those measured for the galaxy distribution. For low density 
models the required amplitude is a better match to the observed
strength of galaxy correlations on large scales. Biasing must 
therefore enhance the contrast of structure in the 
galaxy distribution for the $\tau$CDM and SCDM models (see, for  
example, fig.~16 of Davis et al 1985) whereas for the other models it
is not required. Fig.~1 suggests,
independent of this, that more fine-scale structure is to be expected 
in high density models. The difference between the two high density
models gives a visual impression of the ``lack of large-scale power''
which has often been cited as ruling out standard CDM; large-scale
correlations are consistent with those measured for galaxies and
galaxy clusters in the $\tau$CDM model, but are too weak in SCDM.

\begin{figure}[t] 
\vspace{14.9cm}
\caption{Slices through simulations of four cosmological
models at $z=1$.
}
\end{figure}

\begin{figure}[t] 
\vspace{14.9cm}
\caption{Slices through simulations of four cosmological
models at $z=3$. 
}
\end{figure}

The differences in evolution between the various models are quite 
striking at higher redshift. By a redshift of three 
the $\Omega=1$ models look much more uniform, the $\Lambda$CDM
model has changed rather little, and the OCDM model has hardly
changed at all. These differences reflect, of course, the
different behaviours of the linear growth factor.
In the open case growth effectively ``switches off'' at
$1+z \sim \Omega_0^{-1} \sim 5$. In the $\Lambda$CDM case
this switch-off occurs at $1+z \sim \Omega_0^{-1/3} \sim 1.5$, while
for an Einstein-de Sitter universe growth continues until $z=0$.
Thus in low density models we expect much more large-scale structure
in the high redshift mass distribution than if $\Omega=1$. Because
of the bias galaxy clustering is expected to evolve less
rapidly,
but, as I discuss next, the galaxies themselves 
evolve more rapidly in this case. It is interesting to note
that the pattern of the final large-scale structure is still visible 
at $z=3$ in the high density models and would be more
prominent in the biased but observable galaxy distribution.

\section{Bias and its evolution with redshift}

Except in the few situations where its gravitational lensing
effects can be measured directly, the large-scale structure seen in
figs 1 -- 3 must be investigated using ``tracers'' like galaxies,
galaxy clusters, quasars, or the gas seen as quasar absorption lines.
The simplest method is to map out the spatial distribution of the
tracer, to characterise its properties by some appropriate statistics,
and then to use a model to relate the statistics to those of the 
dark matter. Although direct measurements of peculiar velocities for 
nearby objects are good enough to map the local mass distribution (Dekel
1995; Strauss and Willick 1996), this is impossible at higher redshift
(except that peculiar velocities for galaxy clusters may be measurable 
using the 
kinematic Sunyaev-Zel'dovich effect, e.g. Haehnelt and Tegmark 1996). 
Peculiar velocities can be measured statistically at high redshift
through the anisotropies they induce in the apparent spatial 
clustering of galaxies. Here, however, as with all
clustering statistics, the interpretation hinges critically on the 
relation between the  tracer and the mass distributions, in other 
words on the ``bias''.

In hierarchical clustering theories a good model for the bias of
galaxy clusters can derived from the gaussian initial
conditions (Kaiser 1984). The current amplitude of superclustering 
depends on cluster abundance and on the shape and amplitude of the 
linear power spectrum of mass fluctuations; it has no direct
dependence on $\Omega$ and $\Lambda$ (Mo et al 1996). The evolution 
of superclustering {\it does} depend strongly on the cosmological 
parameters because they change the history of the linear growth 
factor and so the amplitude of linear fluctuations at high $z$.
To apply this test to a sample of distant clusters one would need to
know only: (a) the abundance of the sample; (b) that it is effectively
complete for all clusters more massive than some (possibly unknown)
threshold; and (c) the amplitude of cluster-cluster correlations.

For more abundant tracers like galaxies or absorbing gas, much more
physics must be included to get a realistic model for bias. In
hierarchical theories residual gas is supposed to collapse
dissipatively within the halos provided by the dark matter, settling
to form centrifugally supported star-forming disks at their centres
(White and Rees 1978; Fall and Efstathiou 1980). Recent work has shown
that such a model,
supplemented by Toomre's (1976) idea that ellipticals and bulges form 
by the merging of early stellar disks, can account qualitatively (and
often quantitatively) for most of the systematics of the observed 
galaxy population; e.g. the present distributions of
luminosity, colour, and morphology, and their correlation
with environment (Kauffmann et al 1993), the counts, redshift
distributions and morphologies of distant galaxies (Cole et al 1994; 
Kauffmann et al 1994; Heyl et al 1995; Baugh et al 1996a), the 
observed evolution of the population in rich clusters (Kauffmann 1995,
1996a; Baugh et al 1996b), and the star formation history of disk 
galaxies as inferred from nearby spirals and from the damped
Ly$\alpha$ aborbers in quasar spectra (Kauffmann 1996b). 
This ``semi-analytic'' approach uses simplified but physically based
models to treat each of the important processes (cooling, star formation,
feedback of energy and of metals, evolution of the stellar
populations, rates for galaxy merging). Its results can be compared 
with a much broader range of data than any feasible simulation. The 
main current difficulty, visible in most of the papers cited above, 
is a substantial overprediction of the number of faint galaxies in the local
Universe. Much of the assembly and star-formation of
galaxies is predicted to take place late (at or below $z\sim 1$) if
$\Omega=1$, suggesting rapid evolution of the tracers of large-scale 
structure; earlier formation is possible in low density universes.

Direct simulations which include a dissipative gas component have
confirmed (or inspired) several aspects of the above work, showing
that gas does cool off to make centrifugally supported disks at
the centres of dark matter halos, and that these can plausibly be
identified as the progenitors of galaxies and galaxy clusters (Cen and
Ostriker 1992; Katz et al 1992; Evrard et al 1994). It is currently
impossible to simulate the formation of individual galaxies in
regions large enough to study large-scale structure, and the 
differing compromises with numerical limitations made by different 
groups show up as substantial discrepancies in their predictions 
for galaxy masses and sizes, for the fraction of gas turned into 
galaxies, etc. Although the results of these simulations are
encouraging, none of their
quantitative predictions for ``galaxy'' clustering can yet be
considered reliable. The situation may be better in the case of quasar 
absorbers. Simulations which include both the dark matter and a 
dissipative gas component subject to a photoionising UV background 
seem to give considerable insight into the nature of the absorbers and
into their spatial distribution. It appears relatively easy to explain
both the observed abundance as a function of HI column 
density and the observed coincidence rate between
neighboring lines-of-sight (Cen et al 1994; Hernquist et al 1996; Katz
et al 1996). It is nevertheless still too soon to conclude that the
simulations have converged to the physically correct answer.

The most promising approach to understanding galaxy bias and its
evolution may be a combination of the semi-analytic galaxy
formation models either with similar semi-analytic models for 
clustering (e.g. Mo and White 1996) or with N-body simulations which
do not explicitly follow the gas. A first attempt at each of these routes
was made by Kauffmann et al (1996). This paper shows how the bias in
the present galaxy distribution can be calculated as a function of
galaxy luminosity, colour, or morphological type, as well as how the
consequences of bias for any particular statistic can be evaluated using
N-body simulations. Extensions of this work to larger simulations
such as those shown in figs 1 -- 3 should allow much more detailed
predictions for the evolution of large-scale structure in the galaxy
distribution. These can then be compared directly with the kinds of
data reviewed in the next section.

\section{Measuring large-scale structure at high redshift}

\subsection{Angular correlations}

As shown most recently by the Hubble Deep Field, at faint magnitude
limits the sky is covered with galaxies. With ground-based telescopes
it is possible to get magnitudes, positions and colours for objects
as faint as $B\sim 27$ whereas spectroscopy, even on 10m telescopes
is limited to $B\sim 24$. The faintest galaxies seen are plausibly
(although not necessarily!) the most distant, although the colours of
those found in the HDF suggest, somewhat surprisingly, that only a few
percent are at redshifts beyond 2.5 (Madau et al 1996; Lanzetta et al 
1996; Steidel et al 1996). This appears to require most of the 
star formation and assembly of present-day galaxies to take place {\it
after} $z=2.5$. For these faint samples almost the only available
clustering information is the angular two-point correlation function.
This may be written as:
\begin{eqnarray}
  w(\theta) = A_\gamma\int_0^\infty dz \Bigl({1\over N}{dN\over dz}\Bigr)^2
\xi(\theta d_A,z) {H(z)d_A(z)(1+z)\over c}
\end{eqnarray}
where $A_\gamma$ is a numerical constant which depends weakly on the
slope of the correlation function, $N(z)$ is the number of objects per
unit redshift, $\xi(r,z)$ is their spatial two-point correlation,
$d_A(z)$ is the usual angular size distance,
and $H(z)$ is the Hubble ratio. It is clear that three factors
contribute to the observed $w(\theta)$: the evolution of the galaxies
themselves affects $N(z)$; the evolution of their clustering affects
$\xi(r,z)$; and the background cosmological model affects $H$ and
$d_A$. Notice that $N(z)$ and $\xi(r,z)$ depend on the
precise magnitude and colour criteria used to define the sample since
galaxy abundances and clustering amplitudes depend sensitively 
on luminosity and
colour. Notice also that intrinisically different galaxy populations
will contribute to the integral at different redshifts.

Current data show a steady weakening of $w(\theta)$ as fainter and
fainter samples are considered (Brainerd et al 1995; Villumsen et al
1996). For $R>25$ significant correlations have so far only been
detected for $\theta\leq 1$ arcmin. This corresponds to scales well
below 1 Mpc and so is not really what is normally thought of as 
large-scale structure. Comparison with data at $z=0$ requires
consideration of the evolution of clustering in the strongly {\it
non}linear regime. The current results could be substantially improved
by constructing good deep photometric samples over large fields. A
particularly interesting possibility would be the use of colour
criteria to isolate high redshift subsamples. This should be possible
with the newest wide-field imagers on big telescopes. Preliminary
studies of the dependence of $w(\theta)$ on colour selection criteria
already show a strong, and as yet poorly understood effect (Landy et
al 1996).
 
\subsection{Deep redshift surveys}

Recent deep redshift surveys include the Canada-France Redshift Survey
discussed in this volume by F.~Hammer and O.~LeF\`evre, the 
Anglo-Australian B-limited surveys discussed here by R.~Ellis and 
M.~Colless, and the Hawaii deep survey (Cowie et al 1996). The CFRS,
for example,
contains redshifts for almost 600 galaxies and is about 85\% complete to
a magnitude of $I=22.5$. Its median redshift is greater than 0.5. 
As samples get deeper it becomes {\it much} harder to analyse them
in an analogous way to local surveys. This is not merely because it is more
difficult to get redshifts for individual galaxies, but also
because the sampling volume has a very large
extent in the redshift direction ($\sim 10^3h^{-1}$Mpc) so that many
redshifts have to be obtained before there are enough galaxies within
any given structure (of size 20 to 50 $h^{-1}$Mpc) for it
to be mapped clearly. The situation is made worse by the
expected weakening of large-scale structure with increasing redshift,
and, at very high redshift, by the fact that only a few percent
of faint galaxies are at $z>2$. 

As can be seen from O.~LeF\`evre's contribution, considerations of
this kind led the VIRMOS project to conclude that they need 
redshifts for $10^5$ objects. The use of colour criteria could provide
a well defined sample of preselected high redshift objects to give
this kind of project a better lever-arm for studying the evolution of
clustering. In practice, studies of large-scale structure at high
redshift are likely to be restricted, at least initially, to measuring
two-point correlation functions for galaxies on scales of a few Mpc.
The considerations of previous sections suggest that the major
difficulty in interpreting the results will lie in understanding the
``bias'' of the particular galaxy population observed. Indeed, this is
already true for the CFRS where the major uncertainty in interpreting
the measured correlations at $z=0.5$ is in knowing which population
of galaxies they should be compared with at $z=0$ (LeF\`evre et al 1996).
It is unlikely that the problem of understanding the development of
large-scale structure can be decoupled from that of understanding
galaxy evolution.

Other approaches to structure at high redshift could involve samples
of galaxy clusters or of quasars. Selecting clusters from optical data
is much more complex than, say, measuring a correlation
function, and it is particularly hard at high redshift because of the
large number of foreground and background galaxies. Multiband colours
(``photometric redshifts'') can undoubtedly play a major role in
enhancing the apparent contrast of clusters, and distant cluster selected by
X-ray luminosity or Sunyaev-Zel'dovich decrement may eventually be available. 
Critical points when analysing such samples will be
the influence of the observational selection criteria
and the relation of the distant objects to nearer clusters. For
quasars, of course, similar considerations apply. In both cases the
samples are sparse and so the clustering signal is difficult to
measure. 

\subsection{Large-scale structure in absorption}

There a number of major advantages to using quasar absorption lines to
probe large-scale structure: the probability of detecting any
particular absorber depends only weakly on its position along the
line-of-sight; absorbing systems are abundant along each
line-of-sight; usable lines-of-sight are quite common -- quasars with
$B<20$ have a typical separation of about 15 arcmin; the lines appear 
due to relatively unevolved material so that their relation 
to other components, for example the dark matter, may be relatively
easy to understand; the absorbing gas is plausibly the raw material
for galaxy formation and studies of its distribution and metallicity
should therefore clarify how galaxies form. The best strategy for
carrying out a substantial survey of large-scale structure at $z\sim
2$ to 2.5 is probably to use a multiband, wide-angle photometric sky 
survey to identify quasar candidates; intermediate resolution 
spectroscopy on a 4m-class telescope can then yield a confirmed quasar
sample with a suitable redshift distribution; finally, high
resolution multi-object spectroscopy on an 8m-class telescope would
provide good absorber samples along each line-of-sight. 

The Sloan and 2dF surveys will be able to carry out the first two of
these functions, but they will not have the resolution or sensitivity to see
the abundant, low column density absorbers along lines-of-sight to
quasars with $B\sim 20$. As a result their ability to study clustering
of the absorbers, although very useful, will be limited by the
sparseness of their absorber sample. (For example, the comoving
abundance of detected CIV systems will be comparable to the local
comoving abundance of Abell clusters.) High resolution spectroscopy
on a large telescope is critical to being able to measure structure
reliably on the relatively small scales where it is expected to be
significant at $z\sim 2$.

\subsection{Large-scale structure from gravitational lensing}

Gravitational lensing can be used in at least two different ways
to detect large-scale clustering in the mass distribution. The
first employs the fact that coherent gravitational shearing of the
images of background galaxies induces polarisation, that is to say, 
images which are near each other on the sky have a weak tendency to
line up. This effect can be detected by correlating the orientations 
of galaxies as a function of their angular separation using
large, deep, and high quality photometric images taken during
excellent seeing. This is a difficult measurement
because the gravitationally induced excess ellipticities are of the 
order of one per
cent. So far only upper limits (Brainerd et al 1995) or tentative
detections (Villumsen 1996) of the effect have been published. With
better cameras on large telescopes firm detections are quite
feasible. The measured quantity, the polarisation correlation
function, depends on the redshifts of the background
galaxies, on the amplitude, shape and evolution of the power spectrum
of mass fluctuations, and on the cosmic geometry (see Blandford et al
1991). For geometric reasons most of the effect is induced at $z\sim 0.5$.

A second effect of lensing is produced by its magnification and
demagnification of 
background galaxies. This can result in an apparent clustering
even if the background objects are, in fact, unclustered. The strength
of the effect depends on whether the increased abundance of galaxies in
magnified regions, caused by the lifting of faint systems above the sample
magnitude limit, is outweighed by the increased separation, which 
magnification also produces (Broadhurst et al 1995). 
In practice the combined 
effect is quite weak and must be considered in combination with the 
intrinsic clustering of the faint galaxies. A first theoretical
analysis is given by Villumsen et al (1996). Since image orientations are
not used, there are no additional requirements on image quality beyond
those normally needed to measure angular correlations to faint
magnitude limits. 
%
%

\end{document}